\documentclass[namedreferences]{solarphysics}

\usepackage[hyperref,optionalrh]{spr-sola-addons} 
\usepackage{graphicx}        
\usepackage{color}           
\usepackage{breakurl}       

\chardef\us=`\_

\begin{document}

\begin{article}
\begin{opening}

\title{Solar Radio Imaging at Arecibo: The Brightness Temperature and Magnetic Field of Active Regions}

\author[addressref=aff1,corref,email={mano.rac@gmail.com}]{\inits{P.~K.}\fnm{Periasamy~K.}~\lnm{Manoharan}\orcid{0000-0003-4274-211X}}
\author[addressref={aff1},email={csalter@naic.edu}]{\inits{C.~J.}\fnm{Christopher~J.}~\lnm{Salter}}
\author[addressref={aff2},email={stephen.white.24@spaceforce.mil}]{\inits{S.~M.}\fnm{Stephen~M.}~\lnm{White}}
\author[addressref={aff3},email={phil@naic.edu}]{\inits{P.}\fnm{Phil}~\lnm{Perillat}}
\author[addressref={aff3},email={felix.omar.fernandez@gmail.com}]{\inits{F.}\fnm{Felix}~\lnm{Fernandez}}
\author[addressref={aff1},email={bhakthiperera@gmail.com}]{\inits{B.}\fnm{Ben}~\lnm{Perera}}
\author[addressref={aff3},email={arun@naic.edu}]{\inits{A.}\fnm{Arun}~\lnm{Venkataraman}}
\author[addressref={aff1},email={cbrum@naic.edu}]{\inits{C.}\fnm{Christiano}~\lnm{Brum}}

\address[id=aff1]{Arecibo Observatory, University of Central Florida, Puerto~Rico~00612, USA.}
\address[id=aff2]{Space Vehicles Directorate, Air Force Research Laboratory, Albuquerque, NM, USA.}
\address[id=aff3]{Arecibo Observatory, Yang Enterprises Inc., Arecibo, Puerto Rico 00612, USA}

\runningauthor{Manoharan et al.}
\runningtitle{\textit{Solar Mapping at X-band and Properties of Active Regions} }

\begin{abstract}

Strong solar magnetic fields are the energy source of intense flares and energetic 
coronal mass ejections of space weather importance.  The key issue is the difficulty 
in predicting the occurrence time and location of strong solar eruptions, those 
leading to high impact space weather disturbances at the near-Earth environment. 
Here, we report regular solar mapping made at X-band (8.1 -- 9.2 GHz) with the 
Arecibo 12-m radio telescope. This has demonstrated its potential for identifying 
active regions, about one half to a day in advance, when they rotate on to the 
central meridian of the Sun, 
and predicting the strongest flares and coronal mass ejections directed towards the 
Earth. Results show (i) a good correlation between the temporal evolution  of  brightness 
temperature of active regions and their magnetic configurations; (ii) the ability  
of the mapping data to provide a better picture of the formation sites of active 
regions and to accurately track their evolution across the solar disk, giving forewarning of  
intense solar eruptions leading to severe space weather consequences; (iii) the importance 
of long-term monitoring of the Sun at X-band for understanding the complex three-dimensional 
evolution of solar features as a function of solar activity. 
The key point in this study is the identification of the magnetic properties of active 
regions on the solar disk to aid in improving forecast strategies for
extreme space-weather events.

\end{abstract}


\end{opening}

\section{Introduction}

Solar magnetic fields play a critical role in a wide variety of phenomena occurring on the 
Sun, ranging from slowly evolving structures such as coronal holes, sunspots, and coronal 
loops, to highly dynamical phenomena such as the acceleration of the solar wind, acceleration 
of charged particles, solar flares, and coronal mass ejections (CMEs). 
The magnetic field in the solar atmosphere essentially controls the plasma structure, 
storage of free magnetic energy, and its release at times of flares and/or mass ejections
(e.g., \citealt{fleishman2022}).
The areas of strong magnetic field concentration on the surface of the Sun form active 
regions, which are embedded in groups of sunspots of the same magnetic polarity, followed 
by groups of sunspots of opposite polarity.  Specifically, active regions coupled to  
sunspot groups of complex polarity of $\gamma$ or $\gamma$-$\delta$ configuration, as 
per the Hale or Mount Wilson scheme (e.g., \citealt{hale1919}; \citealt{kunzel1965}), are 
prone to 
produce significantly intense flares and CMEs. However, such spots are limited in number 
to $<$1\% of the total number of spots compared to the numerous $\beta$ spots of 
bipolar characteristic, which produce flares of lower intensity (e.g., 
\citealt{jaeggli2016}). Therefore, the key issue 
is to predict the occurrence time and location of strong solar eruptions, {\it i.e.}, 
those leading to the high impact space weather disturbances in the near-Earth environment.

The solar magnetic fields are observed directly at the photospheric level, whereas direct 
field
measurements are very difficult in the dynamic corona due to its low density. The inference 
of the coronal field is largely via data-driven models, which are limited by the basic 
assumption that the coronal magnetic field 
remains in static equilibrium. However, the X-ray and extreme ultraviolet (EUV) emissions 
from the optically-thin corona above an active region, 
originating at the top of the complex magnetic field network, relate to the inhomogeneous, 
hot, and dense plasma and they provide a remarkable view of the magnetic activity 
above the active region (\citealt{sam2008}). In a close resemblance, but in the extended 
solar atmosphere, the radio opacity decreases with increasing observing frequency 
and the effective radio emission height, from meter to centimeter wavelengths, moves from 
the corona to the chromospheric region. The radio signatures of an active region in 
the frequency range of 5 -- 10 GHz provide a powerful diagnostic of the gyro-synchrotron 
radiation from high-energy electrons trapped in small-scale magnetic field loops and the 
observed bright features are gyro-resonance 
emitting regions where the field strength exceeds 600 G (e.g., \citealt{bastian1998}; 
\citealt{white1999}; \citealt{sbibasaki2011}; \citealt{nindos2020}). 

Typically, the gyro-resonance spectrum peaks in the frequency range of 5 -- 10 GHz 
and provides an {\it indirect} measure of magnetic fields above 
the photosphere (\citealt{gary2018}). Moreover, observations in this frequency range do 
not wholly resemble those at the soft X-ray and/or EUV bands, but they do largely imitate 
the photospheric 
magnetograms (\citealt{dabrowski2009}; \citealt{nita2004}; \citealt{white2011}). Since 
the magnetic complexity of an active region crucially determines the occurrence of 
intense flares and energetic CMEs (\citealt{priest2002}; \citealt{yashiro2005}), the
present study emphasizes the 
importance of regular mapping of the Sun at $\sim$8.1 -- 9.2 GHz in revealing the 
magnetic characteristics of eruptive regions analogous to the magnetogram data. 

In Section 2 of this paper, we provide a brief description of the Arecibo 12-m radio 
telescope. In the following sections, we discuss our X-band (8.1 -- 9.2 GHz) solar 
mapping observations and report significant results on, (i) tracking the 
formation and evolution of active regions leading to strong solar eruptions of space 
weather importance, (ii) understanding the relationship between the radio emission 
brightness and the magnetic field properties of the quiet Sun and flaring regions, 
(iii) the global view of the solar features over several solar rotations, and (iv) 
the ``solar flux density -- brightness temperature'' relationship. Finally we give 
the concluding remarks in Section 8.

\begin{figure}[h]
  \centerline{\includegraphics[width=9cm]{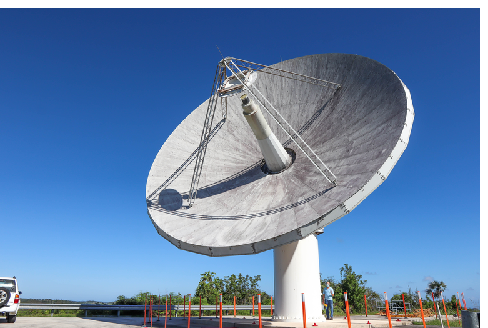} }
  \caption{The hill-top view of the Arecibo 12-m Radio Telescope.}
  \label{fig_1}
\end{figure}

\begin{table}
  \caption{Antenna Location and Parameters}
  \begin{tabular}{ l  l }
  \hline
  Site location & \\
   -- Longitude             & 66$^\circ$ 45$'$ W \\
   -- Latitude              & 18$^\circ$ 20$'$ N \\
   -- Elevation (above MSL) &  496 m \\
  Cassegrain focus, F/D ratio (primary surface) & 0.375 \\
  Surface accuracy (includes fabrication accuracy & \\ 
  and effects of wind, gravity, and temperature) & 0.38 mm (RMS) \\
  Possible frequency range & 2.1 -- 32 GHz  \\
  Pointing accuracy (at wind speed $\lesssim$50 km per hour) & $\sim$30 arcsec \\
  Elevation range & +5 to 88 deg  \\
  Azimuth range   & -180 to 360 deg  \\
  Slew and scan rates & \\
   -- Azimuth & up to $\sim$5 deg/s \\
   -- Elevation & up to $\sim$1 deg/s \\
\hline
\end{tabular}
\end{table}


\begin{table}[h]
  \caption{System Performance}
  \begin{tabular}{l l}
  \hline
   System parameters     &  X-band             \\
  \hline
   Frequency range       &  8.1 -- 9.2 GHz     \\
   HPBW$^@$              &  $\sim$10 arcmin    \\
   T$_{\rm sys}$$^*$     &  123 K              \\
   SEFD$^*$              &  4200 Jy            \\
  \hline
  $^@$HPBW is the telescope$'$s half-power beam width  &  \\
  $^*$T$_{\rm sys}$ and SEFD vary with elevation (see Figure \ref{fig_2}) &  \\
\end{tabular}
\end{table}

\begin{figure}[h]
\centerline{\includegraphics[height=0.5\textwidth]{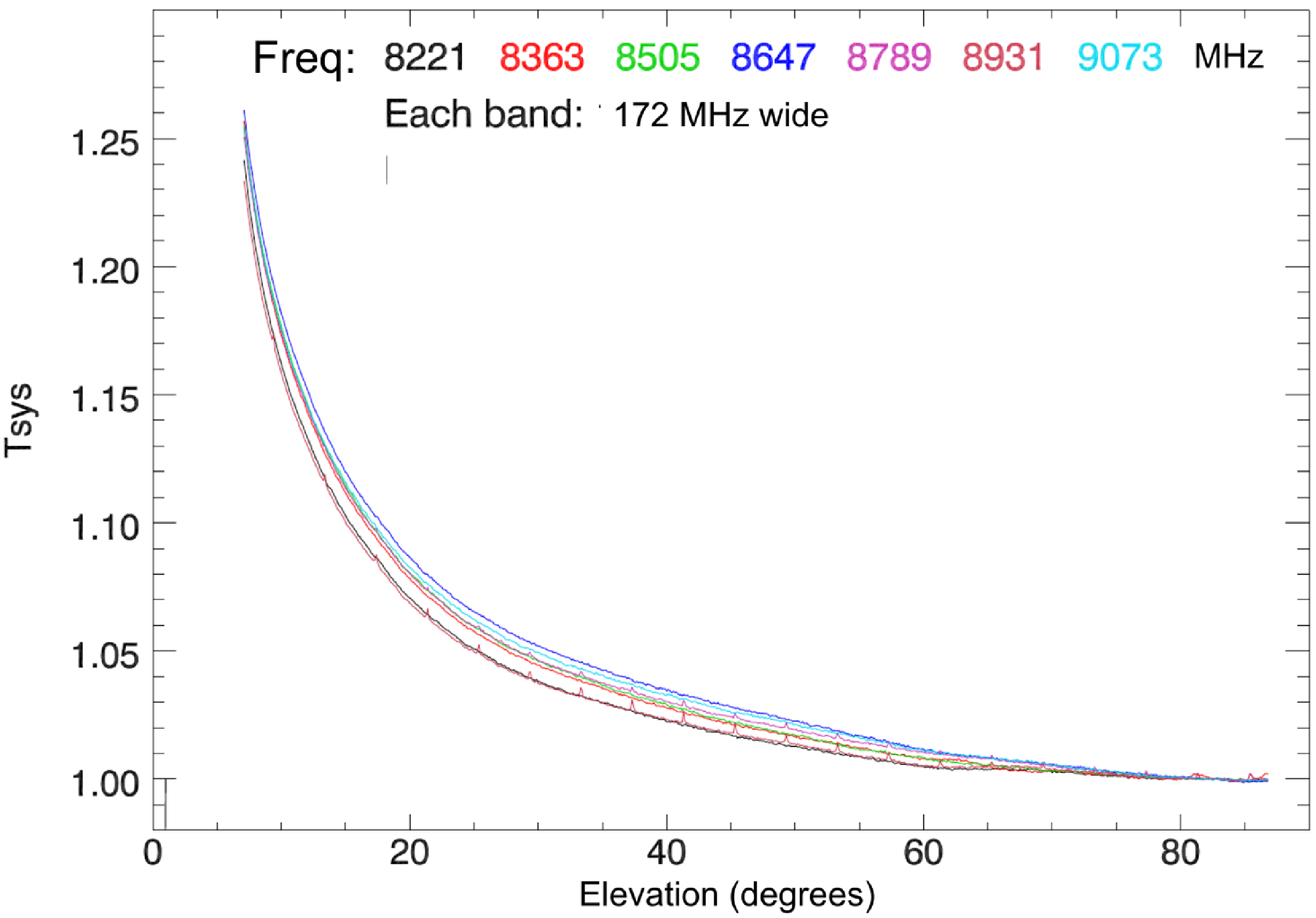} }
\caption{The variation of the 12-m radio telescope's system temperature (T$_{\rm sys}$) 
as a function of elevation angle at X-band.
Seven frequency bands, each of 172-MHz bandwidth, are shown in different colors. Each 
band's T$_{\rm sys}$ is normalized to its average value of 123~K at elevations of 
$\sim$80-85 degrees (see Table 2).
}
  \label{fig_2}
\end{figure}

\section{The Arecibo 12-m Radio Telescope}

The Arecibo Observatory operates a 12-m diameter parabolic reflector radio telescope. 
It was manufactured by Patriot Antennas Inc, Albion, Michigan, USA. 
The 12-m telescope was installed in 2011 on a hill-top within the observatory campus.  
It is a fully steerable alt-azimuth-mount telescope with a primary focal length to 
diameter ratio of 0.375. The telescope can cover an elevation range of 
$\sim$5$^\circ$ -- 88$^\circ$ and provides a coverage in declination, between 
$\sim-$65$^\circ$ and +90$^\circ$. The hill-top view of the 12-m telescope is shown in 
Figure \ref{fig_1}. The geographic coordinates of the telescope site and the mechanical 
specifications of the antenna are given in Table 1.

\subsection{The Room-temperature Receiver Systems}

The 12-m radio telescope operated with room-temperature receiver systems until 10 April 
2023, which covered the frequency ranges of 2.21 -- 2.34 GHz (S-band) and 8.1 -- 9.2 GHz 
(X-band) and recorded dual polarization signals using the FPGA-based Mock Spectrometer,
which contains seven boxes and each box handles bandwidth up to 172 MHz
(\url{http://www.naic.edu/~astro/mock.shtml}).  The 12-m telescope takes advantage of 
RFI protection from the Puerto Rico Coordination Zone (PRCZ) at frequencies below 15 GHz, 
which covers Puerto Rico and nearby Puerto Rican islands.
(\url{https://www.naic.edu/ao/scientist-user-portal/science-tools/pr-coordination-zone}).
As a stand-alone telescope, the observing time of the 12-m antenna is mostly 
shared among the mapping of the Sun at the X-band (e.g., \citealt{mano2022arXiv221104472M};
\citealt{mano2023}) and the monitoring of pulsars and FRBs at the S-band (e.g., 
\citealt{ben2022}; \citealt{ben2023}). 
The observing programs 
of the telescope also include mapping of large angular-size continuum radio sources, 
monitoring of selected AGNs and blazars, spectral line studies, etc.  Additionally, the 
12-m telescope provides strong support for  student programs, such as the NSF-funded 
{\it Research Experience for Undergraduate} (REU) and {\it Partnerships in Astronomy 
and Astrophysics Research and Education} (PAARE) programs.

The properties of the X-band room-temperature receiver are given in Table 2. 
The table lists the Frequency range, Half Power Beam Width (HPBW), the typical System 
Temperature (T$_{\rm sys}$), and System Equivalent Flux Density (SEFD). It is to be noted 
that the system temperature (T$_{\rm sys}$) varies as a function of elevation angle as 
shown in Figure \ref{fig_2}, which shows each band$'$s T$_{\rm sys}$ normalized to its 
average value of 123 K at elevations of $\sim$80-85 degrees.  Its average functional 
form at X-band is given by,
   \begin{equation}
     T_{\rm sys} = 0.9523 + 0.0477 \cdot \sin(E)^{-0.85},
   \end{equation}
where E is the elevation angle.  Correspondingly, the SEFD also varies as a function of 
elevation angle. The X-band, 8.1 -- 9.2 GHz, is divided into seven bands of width 172 MHz,
each covering a box of the Mock spectrometer.
In the case of solar mapping, the system is calibrated before and after each map, by 
taking `on -- off' cal (i.e., a value of $\sim$30~K) at a position away from the Sun.
Since even the quiet Sun is strong at X-band (refer to Figures \ref{fig_15} and 
\ref{fig_16}), the solar attenuator is needed and the uncertainty in the measured 
brightness temperature is rather small. However, the important point is that the 
performance of the X-band system can be considerably affected (i.e., attenuated) by 
rain (i.e., T$_{\rm sys}$ increases with the presence of rain, or even thick cloud 
coverage, along the line of sight to the radio source) and such observations are not 
considered in the analysis. More extensive details of performance of the 
system can be seen  at \url{http://www.naic.edu/~phil/hardware/12meter/sysperf}.

\begin{figure}[h]
  \centerline{\hspace{0.0\textwidth}
               \includegraphics[width=0.9\textwidth]{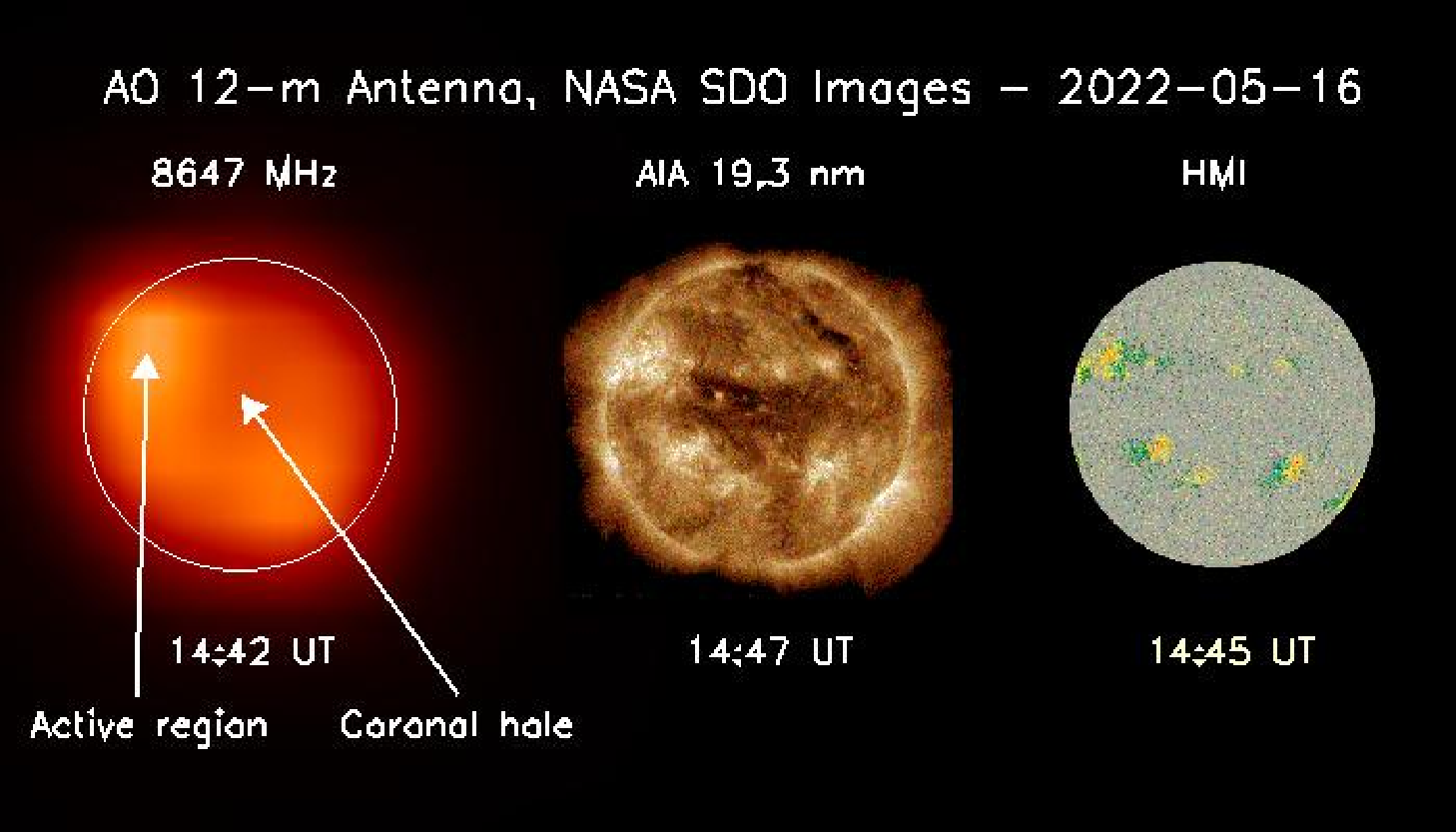} 
              }
  \vspace{0.001\textwidth}
  \centerline{\hspace{0.0\textwidth}
               \includegraphics[width=0.9\textwidth]{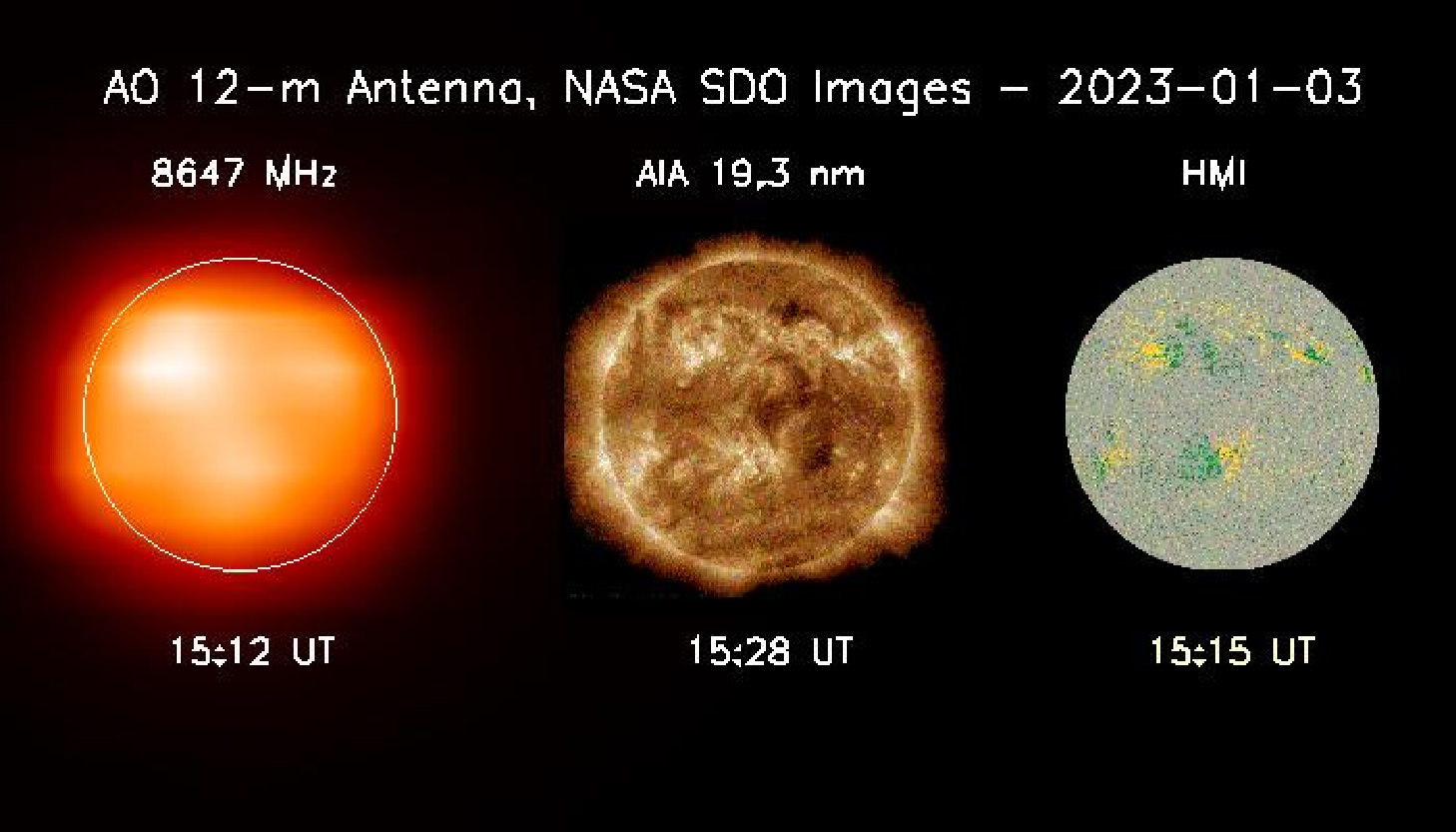}
              }
\caption{(Left images on the top and bottom rows) Sample images of the Sun on 16 May 2022 
and 03 January 2023 at 8647 MHz made with the 12-m Arecibo telescope.  The
white circles on the images indicate the optical disk of the Sun. In the radio image of
the top plot, a bright emitting active region and a low-emitting mid-latitude coronal hole 
are evident and are indicated by arrows. The bottom radio image shows three active 
regions, one located in the southern hemisphere, the other two in the northern hemisphere.  
Each radio image is compared (middle) with the near-simultaneous EUV image of the Sun at 
19.3 nm and (right) the photospheric magnetogram respectively by AIA and HMI on board the 
NASA SDO space mission.
}
  \label{fig_3}
\end{figure}

\section{Solar Radio Mapping}

Solar mapping with the Arecibo 12-m Radio Telescope was initiated in 
mid-December 2021. `East-west' raster scans of the Sun were taken covering 
a range of ${\pm 1^\circ}$ respectively in right ascension and declination, with 
respect to the center of the Sun. Each set of scans provides a 
calibrated map of the two-dimensional distribution of brightness temperature 
over the Sun.  Figure \ref{fig_3} shows examples of full images of the Sun at 
8647 MHz, a frequency close to the center of the recorded band of 8.1 -- 9.2 GHz, 
observed on 16 May 2022 and 03 
January 2023, along with the near-simultaneous EUV images of the Sun observed by the 
{\it Atmospheric Imaging Assembly} (AIA) on board the {\it Solar Dynamics Observatory} 
(SDO) in the wavelength band of 19.3 nm, and the photospheric magnetograms recorded by 
the {\it Helioseismic Magnetic Imager} (HMI) on board SDO (\citealt{pesnell2012}; 
\citealt{lemen2012SoPh}). 
The white circle plotted on the radio images indicates the size of the optical disk of 
the Sun. 

In a day, typically 5 to 10 maps are made to monitor the evolution of the brightness 
temperature distribution of the Sun. 
Since the 12-m radio telescope covers the 
frequency band of 8.1 -- 9.2 GHz, at a given time seven simultaneous maps are made at 
frequency intervals of $\sim$172 MHz. Inter-comparison between these maps provides
a useful handle on the identification and elimination of radio frequency interference, 
should this be present.  Each radio map represents the brightness distribution of the Sun,
in a sense, the `average' characteristics of active 
and quiet regions on the Sun over the scan duration of $\sim$30 min. However, the other 
space-based images compared are snapshots of the Sun with shorter exposure times. 

The spatial resolution of the 12-m telescope at 8.1 -- 9.2 GHz is 
limited to $\sim$10 arcmin. Nevertheless our maps provide a clear view of the emission 
brightness temperatures of active and quiet regions on the Sun. For example, in the radio 
map observed on 16 May 2022 (top row, left image; Figure \ref{fig_3}), the presence of
an active region is identified with an enhanced brightness temperature, $\sim$10,500~K, 
whereas quiet regions are at an average temperature of about 8000~K. The mid-latitude coronal
hole (i.e., open magnetic field region of low density), as indicated by the low emitting
region of the EUV image, is also associated with a low-brightness temperature of about 
$\lesssim$6000~K. Likewise the map on 03 January 2023 (bottom row, left image; Figure 
\ref{fig_3}) shows the presence of three active sunspots groups, one of them located in the 
southern hemisphere and the others in the northern, in correspondence with the EUV image and 
the magnetogram. The brightness temperatures of these active regions lie between $\sim$11,000
and 13,000~K.

\begin{figure}
\centerline{\includegraphics[height=14.5cm]{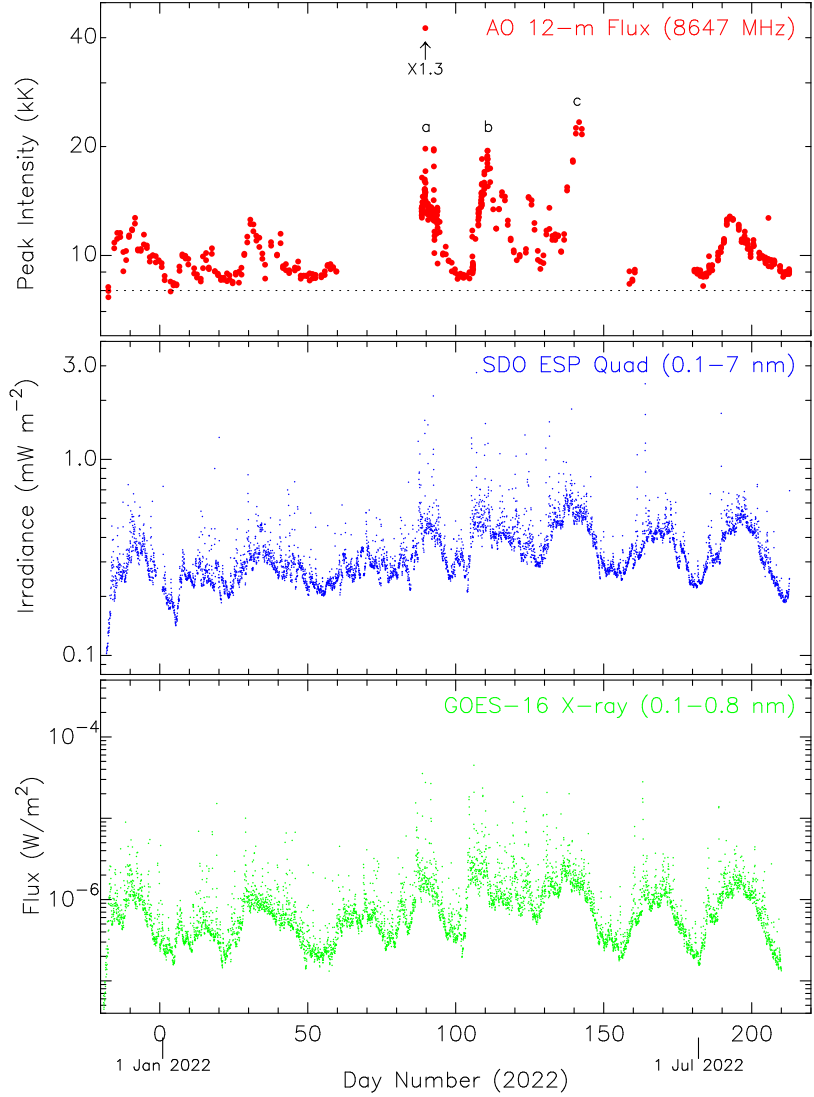} }
\caption{(Top) Peak brightness (in kiloKelvin) of the Sun at 8.6~GHz obtained from the 
Arecibo 12-m radio telescope plotted for dates from 13 December 2021 to 31 July 2022, 
in the ascending phase of the current solar cycle 25. The horizontal dotted represents 
the brightness temperature of the quiet Sun. Data gaps are due to the non-availability 
of the telescope for observation (i.e., taken for maintenance and/or upgrade works).  
(Middle and Bottom) 
Hourly average EUV (EVE/SDO) and X-ray (GOES) fluxes are plotted for comparison. The three 
peaks marked on the radio plot with the letters `$a$', `$b$', and `$c$' are dominant, and 
correspond to strong emission from magnetically-active regions of `$\beta$-$\gamma$-$\delta$'
configuration. }
  \label{fig_4}
\end{figure}

\subsection{Peak Radio Brightness Temperature} 

In Figure \ref{fig_4}, the peak brightness of the Sun (in kiloKelvin), observed with
the 10-arcmin beam of the 8.6-GHz images is displayed in the top panel. This  covers 
the period between 13 December 2021 and 31 July 2022, which is in the onset of the 
ascending phase of the current solar cycle 25. The gradually building up activity of 
the Sun is evident in the brightness, which includes the signatures of active regions 
responsible for X- and M-class flares, as well as quiet days of lesser activity.  In 
the central and bottom panels of the figure respectively, hourly-averaged EUV 
(0.1 -- 7 nm) irradiance of the Sun observed by the {\it Ultraviolet Variability 
Experiment} (EVE) on board SDO (\citealt{wood2012SoPh}) and X-ray (0.1 -- 0.8 nm) 
flux by the NOAA {\it Geostationary Operational Environmental Satellite} (GOES-16) 
(\url{http://www.swpc.noaa.gov/Data/goes.html}) are plotted for comparison. 


In the radio brightness plot, each point indicates the typical average peak brightness on 
the disk of the Sun and each intense peak is associated with an isolated active region on 
the Sun.  In addition, when the mapping time coincided (or partly overlapped) with a flare, 
such data point recorded the brightening corresponding to the particular phase of the flare. 
For example, one of the maps made on day number 89 (30 March 2022) included the rising phase 
of the X-1.3 class flare, the first intense flare of the current solar cycle, and recorded
a brightness temperature of $\sim$42,000~K (see top panel of Figure \ref{fig_4}). Beside 
several systematic peaks, the radio brightness profile reveals the interesting result that 
the 8.6-GHz brightness temperature of the quiet Sun in the absence of activity, is 
$\sim$8000~K, as indicated by the dotted horizontal line in the top panel. This is likely 
the representative temperature of the upper chromosphere of the quiet Sun at $\sim$3000 km 
above the photosphere (e.g., \citealt{zhang1998}; \citealt{avrett2008}).

The strong radio emission from the Sun between day numbers 90 and 140, three peaks 
(indicated by letters `$a$', `$b$', and `$c$' in the top panel of Figure \ref{fig_4})
show systematic increase and decrease that are much more prominent than are seen in 
the EUV and X-ray fluxes. These correspond 
to emission from multiple-pole magnetically-active, `$\beta$-$\gamma$-$\delta$', 
regions developed on the Sun, from where intense flares and energetic CME eruptions 
were observed (e.g., \citealt{jaeggli2016}; \citealt{yashiro2005}), and respectively 
correspond to active regions AR\#2975, ARs\#2993/2994, and AR\#3014. In fact, the 
`$b$'-peak$'$s ARs\#2993/2994 were the return of AR\#2975, which appeared at the east 
limb of the Sun on 15 April 2022, and during the subsequent rotation decayed to a less 
active state.  For each peak, when the 
magnetic configuration of the active region attained $\beta$-$\gamma$ configuration, 
the brightness temperature increased to a level of $\sim$13,000~K. As the peak was 
approached, it reached $\sim$19,000 -- 21,000~K and the magnetic configuration 
developed to $\beta$-$\gamma$-$\delta$. The notable point is that all the 
M-class flares were produced when the brightness temperature was $\geq$13,000~K, 
whereas X-class flares occurred close to the peak of $\sim$20,000~K.

\begin{figure}[h]
  \centerline{\hspace{-0.1\textwidth}\includegraphics[height=0.75\textheight,clip=]{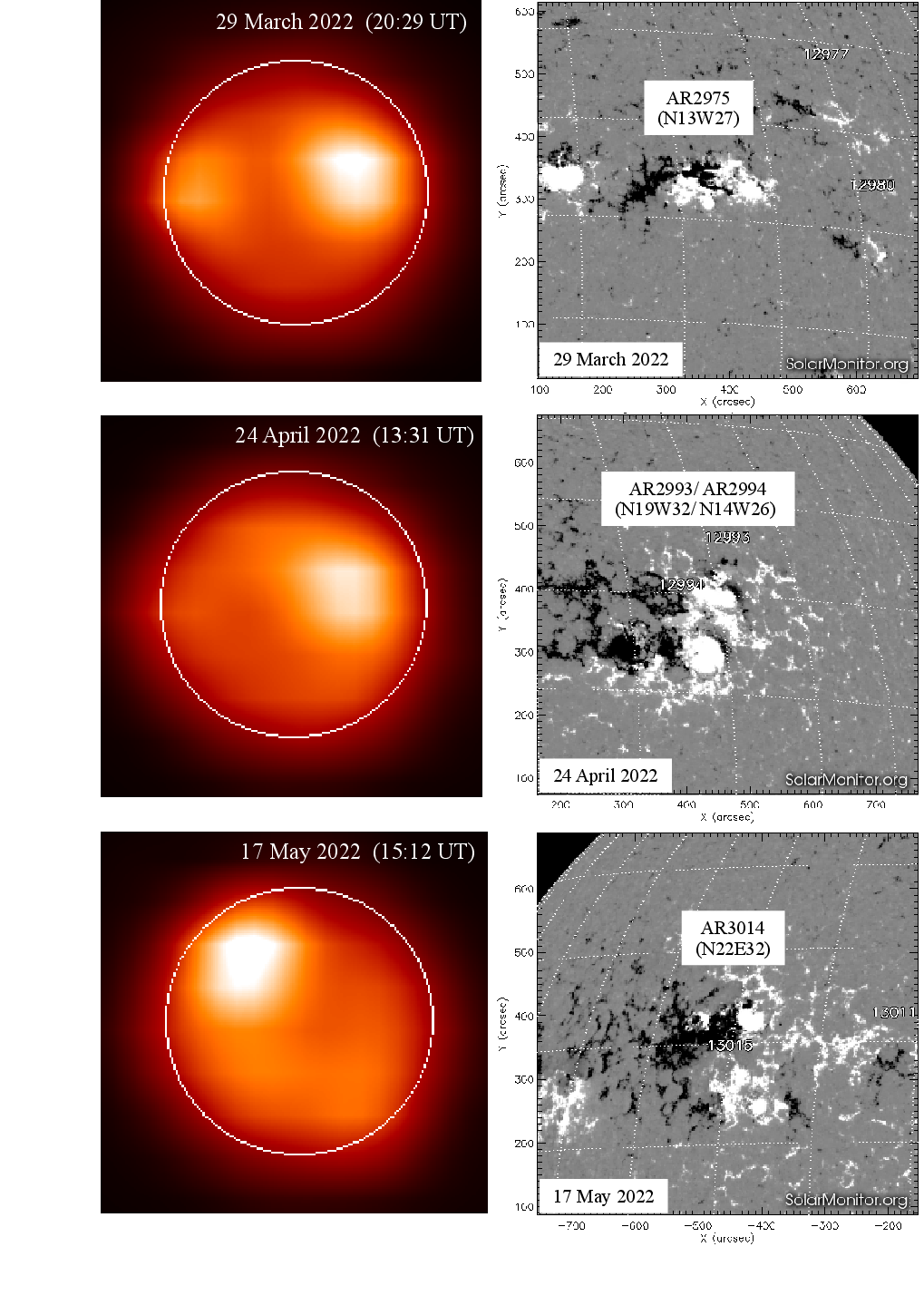}}
     \vspace{-0.05\textwidth}
\caption{(Left) Arecibo radio images at 8.6 GHz from 29 March, 24 April, and 
17 May 2022, with peak brightness temperatures in the range of 13,000 -- 16,000~K and 
of magnetic configurations between $\beta$-$\gamma$ and $\beta$-$\gamma$-$\delta$.  
(Right) The corresponding HMI/SDO magnetograms of the bright emitting portions 
(courtesy of {\it SolarMonitor.org}). }
  \label{fig_5}
\end{figure}

It is valuable to detect rarer intense flares and CMEs, with the help of the 
12-m radio mapping, when an active region attains the $\beta$-$\delta$ 
configuration. Figure \ref{fig_5} shows typical examples of Arecibo radio 
images made at 8.6 GHz on 29 March, 24 April, and 17 May 2022, after the 
development of $\beta$-$\gamma$ magnetic configuration, of peak  brightness 
temperatures in the range of 13,000 -- 16,000~K.  Alongside each radio image
is shown the same day$'$s complex magnetogram of the bright emitting region
from the HMI on board the SDO space mission (\citealt{scherrer2012}).  Such 
an active region while crossing the central meridian of the Sun would release 
Earth-directed CMEs/flares, which are liable to cause severe space-weather 
impacts at the near-Earth space.  However, the occurrence of this type of complex 
active region is infrequent and represented by only $\sim$1\% of the sunspot population 
(e.g., \citealt{jaeggli2016}).  Thus the great advantage of these measurements is 
that they can identify an eruptive region when it rotates close to the central 
meridian of the Sun, about one half to a day in advance, and predict the strongest 
flares and CMEs.


\section{Evolution of X-band Brightness Temperature}

The 12-m radio telescope observations presented in this study 
cover the period between 
13 December 2021 and 09 April 2023, which is in the ascending phase of the 
current solar cycle \#25. The solar mapping observations were terminated at 
$\sim$18~UT on 09 April 2023 and the 12-m telescope was released to the 
telescope engineering team on 10 April 2023 for the 
installation of the wideband cooled receiver system (refer to Section 8). 
The regular radio mapping observations so far accomplished have been 
useful to (i) locate and track several active regions between 
their appearance at the east limb of the Sun and disappearance (i.e., 
rotation to the backside of the Sun) at the west limb, (ii) compare the
evolution of active regions with their daily magnetic properties, such as 
estimated area, number of sunspots within the active group, and magnetic 
configurations, and (iii) study the brightness temperature evolution of the 
Sun as a function of the phase of the current solar cycle \#25.

Several active regions crossed the solar disk without significant 
change, whereas a small fraction evolved from a simple magnetic 
configuration (e.g., $\alpha$ or $\beta$ configuration) to complex 
$\beta$-$\delta$ and/or $\beta$-$\gamma$-$\delta$ configurations, and their 
brightness temperature profiles showed a remarkable increase to a maximum
value, followed by a gradual decrease, or stayed at the increased level 
until they rotated off the visible disk of the Sun.
In Figure 6(a) to (d), daily X-band peak-brightness 
temperature measurements are plotted with red dots for the active 
regions, \#2993, \#3089, \#3153, and \#3190, during their crossing of the 
solar disk.  The cartoons included in Figure 6(c) depict the typical 
locations of the active region \#3153 on the solar disk at three different 
epochs. In the cases of active regions, i.e., \#2993, \#3089, and \#3153, 
during their passage on the solar disk, although each one of them showed an 
increase in brightness temperature to a peak value, followed by a gradual 
decrease, their temperature profiles vary in shape, width and the intensity
during the rotation from the east to the west limb of the Sun. Thus, the
differences in the formation, development, and decay of active regions 
indicate the challenges involved in predicting the intense flare/CME 
space-weather events and emphasize the importance of the regular monitoring 
of active regions, as demonstrated in the current study.

\begin{figure}[h]
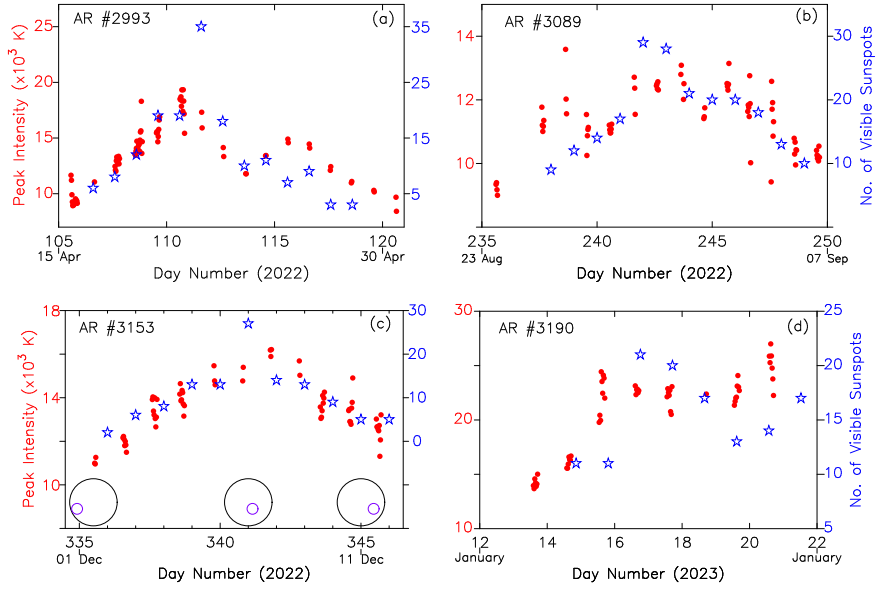

   \centerline{\hspace*{0.00\textwidth}
               \includegraphics[width=3.7cm,angle=-90.0]{figure_6a.ps}
               \hspace*{0.01\textwidth}
               \includegraphics[width=3.7cm,angle=-90.0]{figure_6b.ps}
              }
     \vspace{0.02\textwidth}
            
   \centerline{\hspace*{0.00\textwidth}
               \includegraphics[width=3.7cm,angle=-90.0]{figure_6c.ps}
               \hspace*{0.01\textwidth}
               \includegraphics[width=3.7cm,angle=-90.0]{figure_6d.ps}
              }
   \vspace{0.02\textwidth}

\caption{Plots (a) to (d) show the evolution of peak brightness temperature (in kilo Kelvin) of 
active regions as a function of day number. X-band brightness temperature data points (at 8.6 
GHz) are plotted with red dots and the corresponding vertical axis is shown in red on the 
left-hand side of the plots. Daily estimates of the number of visible sunspots within each active 
region, obtained from NOAA SWPC, are compared with brightness temperature and plotted with blue 
star symbols. Its corresponding scale is shown in blue on the right-hand side vertical axis. 
In plot (c), the solar-disk cartoons depict the typical 
rotation of the AR\#3153 from the east to west limb of the Sun in the southern hemisphere.}
\label{fig_6}
\end{figure}

\begin{figure}[h]
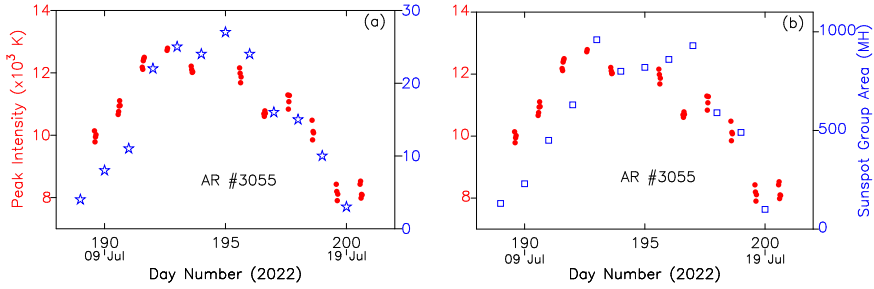


   \centerline{\hspace*{-0.015\textwidth}
               \includegraphics[width=3.7cm,angle=-90.0]{figure_7a.ps}
               \hspace*{0.015\textwidth}
               \includegraphics[width=3.7cm,angle=-90.0]{figure_7b.ps}
              }
   \vspace{0.02\textwidth}

\caption{Plots for the AR~\#3055, similar in format to Figure 6, respectively compare the 
X-band temperature evolution with the visible number of sunspots and the area of the sunspot 
group in millionths of the solar hemisphere (MH) (blue square symbols).  }
  \label{fig_7}
\end{figure}

\subsection{The Magnetic Configuration of Active Regions}

A daily solar active region summary report is compiled and prepared based on the 
analysis of individual observatory reports from the Solar Optical Observing 
Network (SOON) by the NOAA Space Weather Prediction Center (SWPC) and US Air 
Force (USAF) (\url{https://www.swpc.noaa.gov/products/solar-region-summary }). 
The report released at the start of each day typically includes the 
location of an active region on the face of the Sun, its area, the number 
of visible sunspots within 
the region group, and the type of its magnetic configuration. The results of
Figure 6(a) to (d) are compared with the daily magnetic properties of the 
active regions. For instance, close to the maximum of these brightness temperature 
profiles, the magnetic configurations of active regions, \#2993, \#3089, 
\#3153, and \#3190, developed to  $\beta$-$\gamma$, $\beta$-$\delta$-$\gamma$, 
$\beta$-$\gamma$, and $\beta$-$\delta$-$\gamma$, respectively.
The maximum brightness temperatures ranged from $\sim$16,200 to
$\sim$27,000~K, and each active region went through different types of 
evolution.

Additionally in Figure 6(a) to (d), for each active region, along with the 
measurements of the brightness temperatures, the daily estimates of the number 
of visible sunspots within the active region group are plotted with blue
star symbols, with the corresponding scale shown on the right side of the 
vertical axis of each plot. The correlation between the brightness 
temperature of an active region and the development of sunspots (i.e., 
magnetic activity)
is obvious from these plots. However, a comparison between the 
number of sunspots and brightness temperature near the maximum suggests a likely 
range of temperatures for a given number of sunspots, or vice versa.
The results are further discussed in Section 5, which examines the
observations of flare activity in the period between 13 December 2021 
and 09 April 2023 (also refer to Figure \ref{fig_9}).

Figure 7(a) shows the X-band temperature profile of active region \#3055, which 
attained a maximum of $\sim$12800~K, which is less than the maximum values
measured for the other active regions shown in Figure 6.  Moreover, the 
magnetic configuration of this active region also evolved from the $\alpha$ type 
at the east limb of the Sun to the $\beta$ type around the peak of the temperature 
profile, but not to a complex state. It returned again to the initial state close
to the west limb, where the measured temperature was around 8000~K, 
which is in agreement with the long-term quiet Sun (or background) emission 
(see Figure \ref{fig_4}). Similar to the temperature profile, 
the number count of sunspots in the region also evolves (or increases) from a 
count of 4 at the east limb to a maximum of 27 and then decays to a count of 
3 at the west limb. In Figure 7(b), temperature measurements of the active 
region are compared with the area estimates of the active region, given in 
millionth of the solar hemisphere area (MH),
available from the daily solar region summary reports 
(\url{https://www.swpc.noaa.gov/products/solar-region-summary}). The good 
agreement observed between ``brightness temperature and active group area'' 
has also been confirmed for other active regions. In the 
next section, we examine the relationship between the number of sunspots in 
an active group and its area for a large number of flaring sites
located close to the central meridian of the Sun.

\begin{figure}[h]
  \centerline{\hspace{-0.1\textwidth}
              \includegraphics[width=0.4\textwidth,angle=-90]{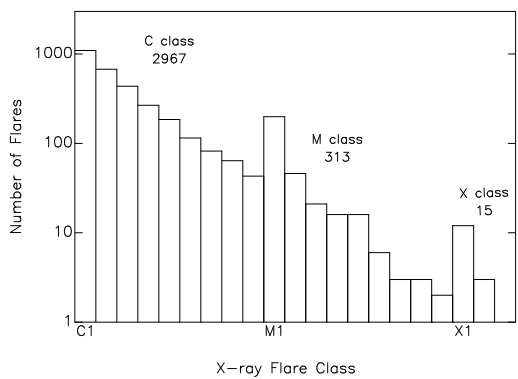}
              }
\caption{Distribution of the number of flares of magnitudes from C1 to X2 classifications 
observed during the period between 13 December 2021 and 10 April 2023. In this ascending phase 
of solar cycle \#25 (see Figure \ref{fig_13}), the number of flares of classifications from 
C1 to X2 is 3295. About 10\% of these are M-class events. The X class flares represent only 
$\sim$0.5\% of the population of the total number of events shown in the figure.
}
\label{fig_8}
\end{figure}

\begin{figure}
  \centerline{\hspace{0.0\textwidth}
              \includegraphics[width=0.4\textwidth,angle=-90]{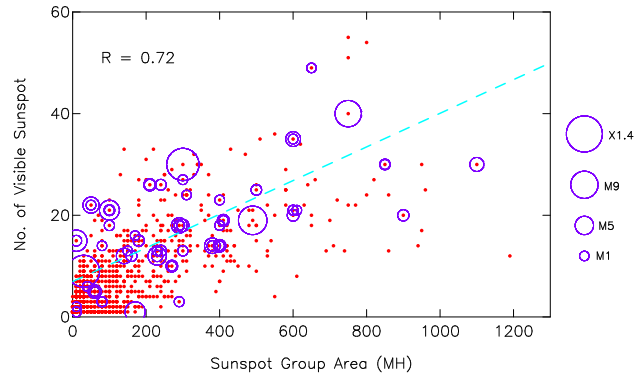}
              }
\caption{Scatter plot between the number of visible sunspots of flare-producing active 
region groups and their associated sunspot group areas. These active regions located within
$\sim$$\pm$30$^\circ$ from the Sun's center produced flares of magnitude $\ge$C1 (see 
Figure~\ref{fig_8}). The number of sunspots shows an increasing trend with the area and the 
linear correlation coefficient is $\sim$0.72. However, for a given area of a group, a 
large scatter is seen in the number of sunspots. The active regions producing flares of M 
and X classes are indicated by circles, and the diameter of the circle corresponding to the 
magnitude of the flare.
}
\label{fig_9}
\end{figure}

\section{The Magnetic Properties of Flare-Producing Active Regions}

In continuation of the above results on the relationship between ``the 
brightness temperature and magnetic property of active region'', we also 
carefully examined the magnetic characteristics of moderate to intense 
solar flare producing regions.  
Flares are classified according to their X-ray intensity in the wavelength 
range 0.1 to 0.8 nm. The weakest flares are A- and B-class, followed by 
C-, M- and X-class, the increase in X-ray intensity from one class to the 
next being by a factor of 10, where the intensity of a C1 flare is
10$^{-6}$ W m$^{-2}$ and the X1 is at 10$^{-4}$ W m$^{-2}$. 
In the current period of study, 13 December 2021 to 09 April 2023, a total 
number of 3295 X-ray flares of intensity C1 and above were observed, their  
distribution being shown in Figure \ref{fig_8}. These flares were associated with 
active regions of different magnetic configurations of less evolved and simple 
$\beta$ types to evolved and complex $\beta$-$\gamma$-$\delta$ stages.  A great 
fraction of them, i.e., 2967 events ($\sim$90\%), are in the weak-to-moderate 
intensity C-class category, 313 events ($\sim$9.5\%) in the 
M-class, and only the 15 remaining events ($\leq$0.5\%) in the intense X1 and 
X2 classes. Thus, in the ascending phase of solar cycle \#25,  
intense flares were limited to a small fraction. 

In the above set of 3295 flare events, 1028 events that originated close to the 
central meridian of the Sun, i.e., within $\pm$30$^\circ$ of the longitude and 
latitude of the Sun's center, have been selected. For each one of these,
the magnetic properties of the flaring site are compared. Figure \ref{fig_9} 
displays the scatter plot between the number of visible sunspots in the flare region 
and its area, which is given in millionth of the solar hemisphere area (MH). 
In spite of the large scatter in the plot, there is an overall increasing 
trend between the number of sunspots and the area, with a correlation 
coefficient of +0.72. The active regions which were responsible
for intense flares of M and X classes are marked with circles. The diameter of
the circle indicates the flare intensity as given in the legend at the right-hand
side of the vertical axis. An observed active region, due to its possibility 
of non-evolving or evolving condition, may be represented in the 
plot by a single point or more than one point. Moreover, on some occasions,
an active region might have been the source of more than one 
flare of the same or varying strength, which are indicated by concentric 
circles on a point.

The above results of magnetic properties of flaring sites indicate that 
for an active region, at the time of its evolution or at the developed complex
state (e.g., $\beta$-$\delta$ or $\beta$-$\gamma$-$\delta$), it is a 
great challenge to accurately pinpoint its relationship to either the 
sunspot count in the group or the area. In contrast, as demonstrated in 
Figures \ref{fig_6} and \ref{fig_7}, the X-band solar mapping likely 
provides a better 
picture of active regions as they evolve and helps to track their
evolution across the solar disk, forewarning of intense solar 
eruptions.

\begin{figure}
  \centerline{\hspace{0.0\textwidth}
               \includegraphics[width=0.47\textwidth,angle=-90]{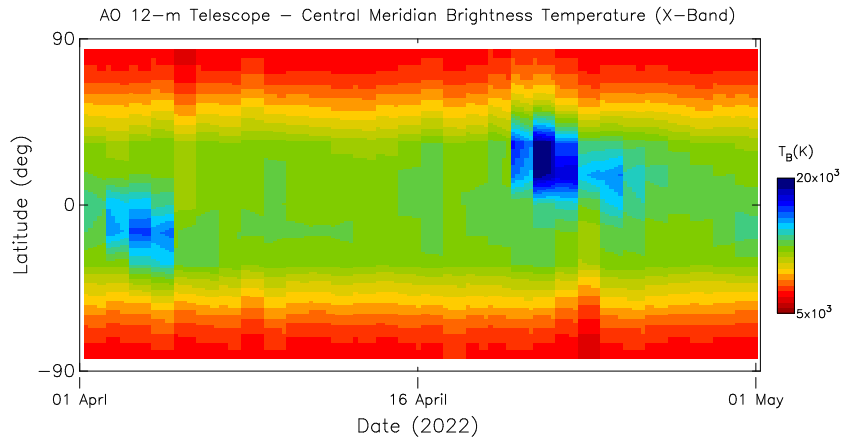}
              }
\caption{A synoptic map made from a sequence of central meridian cuts taken
from the daily images of the 12-m telescope at 8647 MHz between 01 and 30 April 2022. It 
covers the full Carrington rotation \#2256 and part of the end portion of rotation \#2255. 
Two bright active regions centered respectively around 03  and 21 April 2022 are clearly 
seen in the brightness distribution. The sunspot belt of moderate emission occupies the
$\sim$30$^\circ$ of the equatorial region. The stability of the long lived low-emitting 
coronal holes in the high-latitude regions are evident in the map (see Section 6).
}
  \label{fig_10}
\end{figure}


\begin{figure}[h]
  \centerline{\hspace{0.0\textwidth}
               \includegraphics[width=0.47\textwidth,angle=-90]{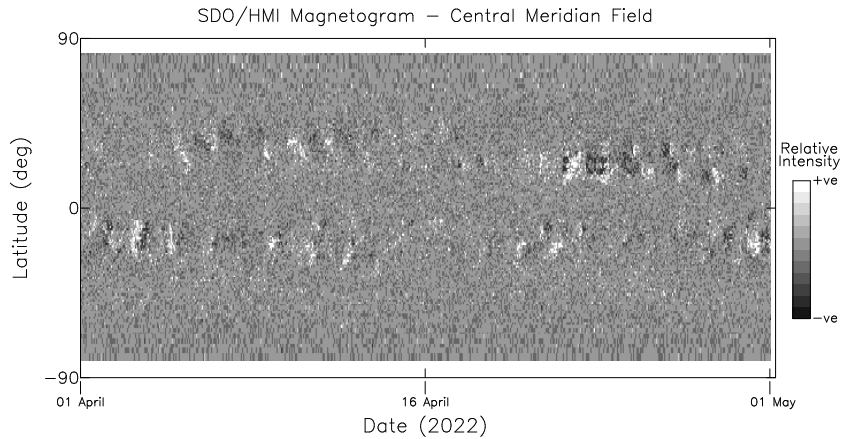}
              }
\caption{The synoptic map of the photospheric magnetic field generated using 
near-central-meridian magnetogram data from the HMI/SDO for the month of April 
2022 is plotted in the same format as Figure \ref{fig_10}. 
}
  \label{fig_11}
\end{figure}

\section{Solar Synoptic Maps} 

\subsection{Brightness Temperature Distribution}

Apart from the daily radio maps of the Sun, displaying its brightness 
temperature distribution over a solar rotation is an efficient way
to represent the full surface features of the Sun during a rotation,
yielding a global view of emission structures. Moreover, it is a valuable 
tool for assessing the conditions of the emission characteristics in 
relationship to the solar magnetic field evolution. Figure \ref{fig_10} 
shows an example of a solar synoptic map, obtained using the continuous data 
from the 12-m telescope at 8647 MHz for the month of April 2022. This
includes the full Carrington 
rotation \#2256, spanning from 03 to 30 April 2022 and end part of the 
previous rotation. To construct the synoptic map, a central meridian cut 
of width $\sim$5~arcmin is taken from the daily radio images, covering
from the south to the north poles. Since the spatial resolution of the 
12-m telescope at X-band is $\sim$10~arcmin, the strip of 
$\sim$5~arcmin at the center of the Sun corresponds to the peak of the 
beam. The vertical central meridian strips are assembled horizontally 
in chronological order and deprojected to get a uniform latitudinal view of
the solar features.

The synoptic map representation allows us to examine the temporal evolution of 
surface characteristics of the Sun. For example, in Figure \ref{fig_10}, two 
bright regions, 
one in the southern hemisphere, centered around 03 April (due to the combination of 
active regions ARs\#2978 and \#2981, both of which developed to the complex 
$\beta$-$\gamma$ configuration), and the other in the northern hemisphere, 
centered around 21 April 2022 (due the active regions ARs\#2993 and \#2994 of 
$\beta$-$\gamma$ configuration) are evidently seen. Although both of them
have a similar sort of magnetic configuration, the later active region's 
brightness temperature was much higher than that of the earlier one. This 
indicates that the small scale magnetic structures favorable in accelerating
particles to high energy are likely present in the bright emitting region
(e.g., \citealt{panka2010}).
Other interesting structures (or features) are the low emitting strips 
(i.e., $<$7000~K) extending from the north and south poles to mid latitudes, 
which are due to the small to large-size transient coronal holes (represented 
by red color code). Examples are at around 05 April and 22 April 2022. 
Consistently, the stability of the long lived low-emitting coronal holes in 
the high-latitude regions are brought out on the map. However, at the south 
and north edges of the map, the effect of the sky dilution can tend to reduce 
the temperature at a constant level and about 7-degree portion of the polar
region edges are avoided while preparing the map. !in the map display. 
%
The bright emitting belt of about $\pm$35$^\circ$ caused by the presence of 
sunspots (represented by the green color code) is apparent in the map. There 
are also a number of dark green features seen on the equatorial belt, 
indicating the brightening of low-latitude sunspots.  Due to the 
limited resolution of the 12-m telescope, the fine structures 
in the map have been smoothed to the beam size. 
Nevertheless, most of the large-scale features
and low- and high-emitting regions are revealed. 


\subsection{The Photospheric Magnetic Field Distribution}

The X-band brightness temperature distribution of the Sun has been compared 
with the magnetogram map of the observed photospheric magnetic field by HMI
on board SDO. Figure \ref{fig_11} shows the synoptic map prepared with the 
near-central-meridian data from magnetograms acquired everyday around at $\sim$12 
UT in the month of April 2022, which corresponds to the ascending phase of the
current cycle \#25 just after its minimum phase. The positive and negative 
polarities are respectively shown in white and black shades. At the initial
phase of a solar cycle, sunspots appear at the high latitudes in the southern 
and northern hemispheres as is clearly revealed by two striking horizontal
strips on the synoptic map. Moreover, the concentrated magnetic fields above two 
prominent active regions, corresponding to the bright emitting regions on the 
radio map (see Figure \ref{fig_10}), can also be easily identified. Such a 
map is extremely useful to study the global distribution of solar features as a 
function of solar rotation.

\begin{figure}[t]
  \centerline{\hspace{0.0\textwidth}
              \includegraphics[width=0.52\textwidth,angle=-90]{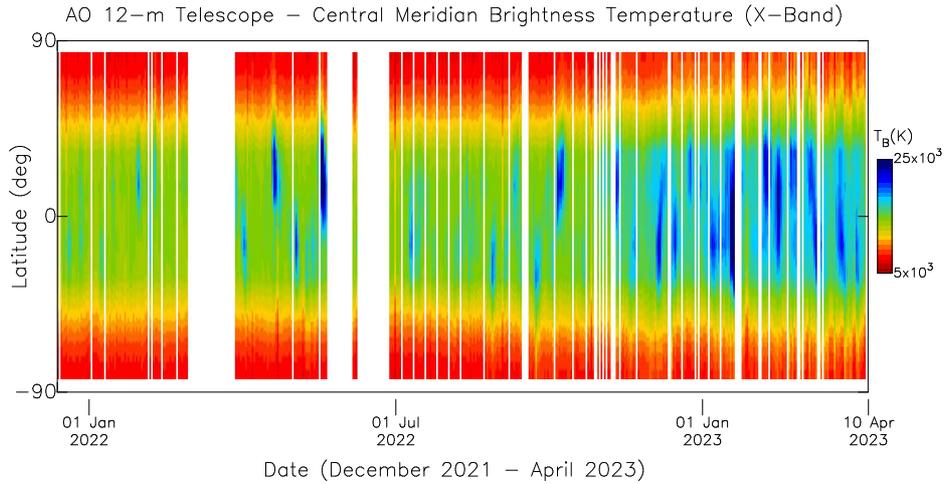}
              }
\caption{The heliographic latitudinal distribution of X-band brightness temperature 
observed by the Arecibo 12-m radio telescope as a function of the date.  In the rising 
phase of the current solar cycle \#25, around January 2022, the brightness around the 
equatorial closed-field region of the Sun was narrow in latitudinal spread compared to 
the brightness distribution observed in March -- April 2023. The white patches represent 
observational gaps due to the telescope maintenance.
}
\label{fig_12}
\end{figure}

\section{Ascending Phase of Solar Cycle \#25}

\subsection{Latitudinal Distribution of the Sun$'$s Brightness Temperature} 

\begin{figure}[t]
  \centerline{\hspace{0.0\textwidth}
              \includegraphics[width=0.65\textwidth,angle=-90]{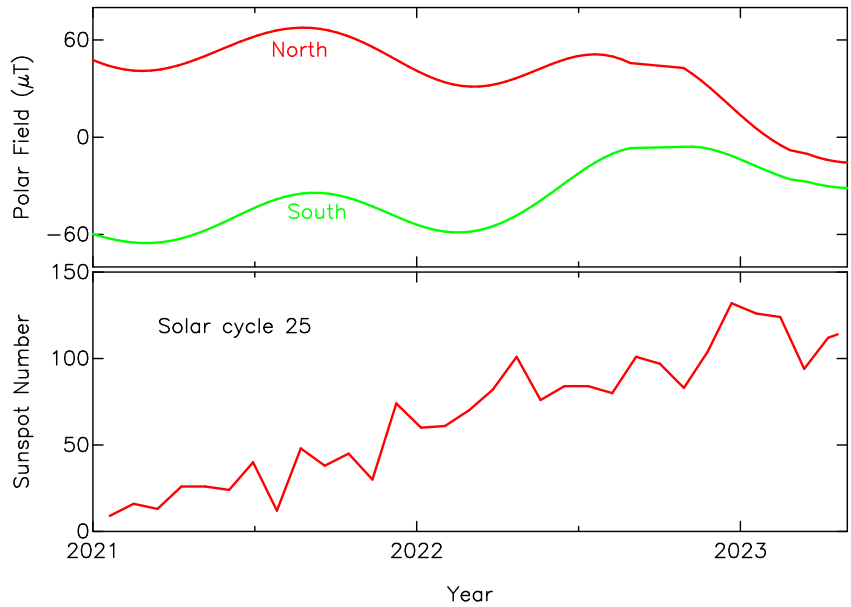}
              }
\caption{(Lower) The sunspot number, and (upper) the intensity of the polar magnetic field 
(red and green lines respectively for north and south polar regions), plotted as a function 
of the year. The increase of solar activity revealed by the sunspot count, as well as the 
gradual decrease of polar field strength, suggests a change from a near dipole magnetic 
configuration of the Sun to a complex structure when the activity increased 
(\url{http://wso.stanford.edu/} and \url{http://omniweb.gsfc.nasa.gov/}).
}
\label{fig_13}
\end{figure}

The 12-m telescope observations covered a good portion of the ascending phase of the current 
solar cycle \#25, between 13 December and 09 April 2023. More than 2000 solar images 
have been obtained. Since the data covered a large number of solar rotations, Carrington 
rotations from \#2252 to \#2268 (plus the beginning portion of rotation \#2269;
i.e., more than 17 rotations), the daily maps include the characteristics of 
repeating active regions, which lived for 
more than one solar rotation. In order to visualize the radio features of the Sun over these
rotations, the same procedure used in preparing the radio synoptic map (see Figure \ref{fig_10}) 
is employed and a central meridian strip of width $\sim$5~arcmin is taken from each day's 
image and the ``{\it Latitude -- Time}'' plot of the brightness temperature is constructed as 
shown in Figure \ref{fig_12}.  This plot includes observations taken during the entire period
of 13 December 2021 -- 09 April 2023.  In a broad sense, the ``{\it Latitude -- Time}'' plot 
resembles the 
``{\it sunspot butterfly}'' diagram and reveals the changes of the major large-scale magnetic 
and brightness structures and the development of activity occurring in the ascending phase of 
solar cycle \#25. Despite several observational gaps, the brightness distribution shows a
number of interesting features and the evidence of a steady increase in solar activity.
For reference, the sunspot number and the strength of the smoothed polar 
magnetic fields are also displayed in Figure \ref{fig_13} with the help of data sets from the 
Wilcox Solar Observatory (\url{http://wso.stanford.edu/}) and the NASA/GSFC’s OMNI database
(\url{http://omniweb.gsfc.nasa.gov/}).

In Figure \ref{fig_12}, the presence of an active region is identified with enhanced 
brightness temperature, $>$8000 K. Some of the active regions of large area (i.e., 
extended in longitude) have taken many days to cross the central meridian line and 
their signatures are wider along the time axis. The quiet Sun regions are at an average 
temperature of about 8000 K.  
In the case of a well-developed active 
region of complex magnetic configuration, a much higher brightness temperature is observed. 
For instance temperatures $\geq$13,000 K serve to identify active regions susceptible to 
intense flares and energetic Earth-directed CMEs leading to severe 
space-weather impacts. The notable point is that all the M-class flares were produced 
when the brightness temperature was $\geq$13,000~K, whereas X-class flares occurred 
close to the peak at $\sim$20,000~K (see Figures \ref{fig_5} and  \ref{fig_6}).

\begin{figure}[t]
  \centerline{\hspace{0.0\textwidth}
              \includegraphics[width=0.52\textwidth,angle=-90]{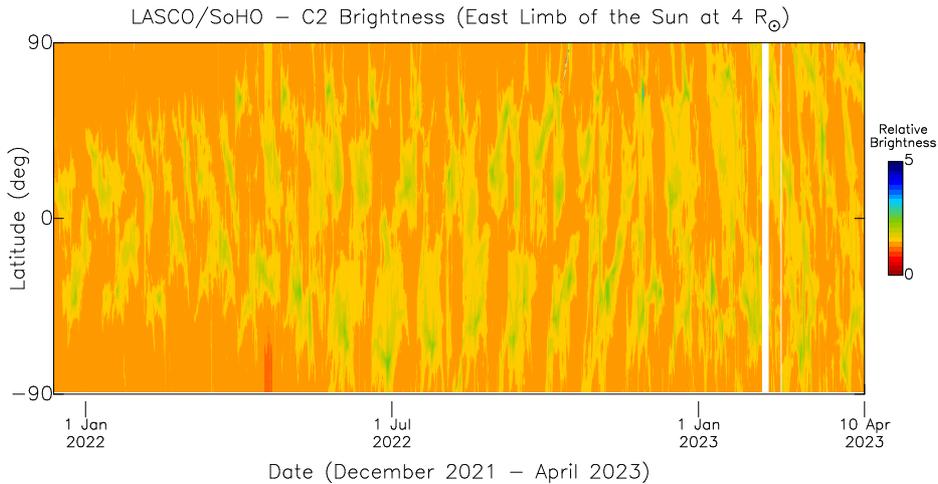}
              }
\caption{Heliographic latitudinal distribution of Thomson-scattered brightness observed at 
4 solar radii by the LASCO-C2 coronagraph, as a function of the date. This plot covers the 
same period as Figure \ref{fig_12}.
}
\label{fig_14}
\end{figure}

In Figure \ref{fig_12}, the increase in the number of active regions in association with the 
sunspot count is shown by the number of bright emitting regions. The development of solar 
activity is also clearly indicated by the gradual increase in the latitudinal width of the 
brightness distribution as the sunspot number increases from mid-December 2021 to April 2023. 
In general, as the solar activity increases, the progressive shrinking of the low-density
as well as low-emitting coronal holes towards the poles is observed (e.g., \citealt{mano2012}; 
\citealt{hathway2015}, \citealt{hamada2020}). In the corresponding phase of solar cycle 
\#25, the brightness temperature distribution is consistent with the structural change of the 
coronal holes in the polar regions respectively shown the gradual decrease in polar field 
strength at latitudes greater than 
60 degrees (Figure \ref{fig_13}) and the density distribution inferred from the
white-light coronagraph image (Figure \ref{fig_14}). 
But, as discussed in Section 6.1, the sky dilution can add a constant reduction to the 
brightness temperature at the south and north edges of the map.
%
%
%
The presence of mid- and low-latitude coronal holes at the central meridian of the Sun, shown 
by X-ray and EUV images of the Sun, is also observed as vertical low-emitting strips on the 
latitudinal distribution of Figure \ref{fig_12}. 
The ``Latitude -- Time'' plot, not only presents a clear picture of
the development of solar activity, but also provides the three-dimensional view of radio 
emission at chromospheric heights in association with the solar surface features.

\subsection{Density Evolution in the Near-Sun Region} 

The above results for brightness temperature distribution observed with the 12-m telescope are 
also consistent with the latitudinal distribution of Thomson-scattered brightness observed by 
the LASCO-C2 coronagraph on board the {\it Solar and Heliospheric Observatory} spacecraft
(e.g., \citealt{brueckner1995}; \url{https://lasco-www.nrl.navy.mil/}).
Figure \ref{fig_14} shows 
the ``{\it Latitude -- Time}'' image of the white light, which is associated with the density of 
free electrons, measured at 4 solar radii above the east limb of the Sun by the LASCO-C2 
coronagraph for the period of the X-band plot shown in Figure \ref{fig_12} 
(e.g., \citealt{mano2009}; \citealt{barbara2011}).
The gradual increase in the latitude width of high 
density features, associated with closed-field active regions, is consistent with the distribution 
of X-band brightness temperature. Moreover, the decrease in the width of the low-density coronal 
hole regions at the poles are consistent with the brightness temperature distribution. 
Another remarkable feature seen in the LASCO plot is the low-density vertical patch near the south 
pole centered around 11 - 15 April 2022. This was associated with a large coronal hole located at the 
south 
pole of the Sun and extended from the pole to a latitude of about 50 degrees. The LASCO-C2 observed 
the low-density feature at the east limb of the Sun. As this coronal hole rotated to the center of 
the Sun in about 7 days, it was recorded by the 12-m telescope around 21 - 22 April 2022 (see
Figure \ref{fig_12}).

\subsection{Radio Flux Density as a Function of Observing Frequency}

The latitudinal manifest of the brightness temperature over about 
17 solar rotations is apparent in Figure \ref{fig_12}. The bright patches are
regions of gyro-resonance (or gyro-synchrotron) emission, high magnetic field 
and significant plasma heating (i.e., due to high number density). This emission 
has a spectral peak in the range of 5 -- 10 GHz (e.g, \citealt{nindos2020}; 
\citealt{white1997}) and it dominates the emission at frequencies, below 3 GHz 
(optically thick part of the solar atmosphere).  It will be useful to compare the 
X-band temperature distribution with the time series analysis of radio flux 
measurements of the Sun at different wavelengths (e.g., in the meter to centimeter
wavelength range).  Such measurements include (i) background emission
from the quiet Sun which does not vary with time (e.g., \citealt{kundu1965})),
(ii) the slowly varying component, caused by free-free thermal emission 
(e.g., \citealt{white1997}), and 
(iii) transient or sporadic emission caused by non-thermal electrons 
accelerated at flare and/or CME sites (e.g., \citealt{mano2003}).

\begin{figure}
  \centerline{\hspace{0.0\textwidth}
              \includegraphics[width=0.55\textwidth,angle=-90]{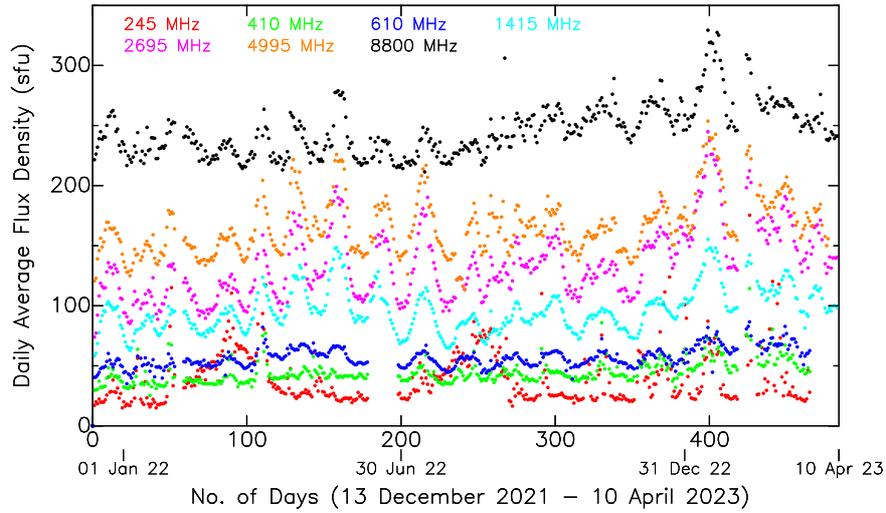}
             }
\caption{Daily average solar flux densities for the period between 13 December 2021 and 10
April 2023 at frequencies of 245, 410, 610, 1415, 2695, 4995, and 8800 MHz. The data sets are from
the Learmonth Solar Observatory, which is one of the stations of the USAF Radio Solar Telescope 
Network (RSTN) (\url{https://www.sws.bom.gov.au/WDC/}).
}
\label{fig_15}
\end{figure}

Figure \ref{fig_15} shows the daily average solar flux densities observed 
in the frequency range of 245 -- 8800 MHz 
by the Learmonth Solar Observatory, which is one of the stations of the 
Radio Solar Telescope Network (RSTN).
The solar flux density is normally 
expressed in solar flux units (sfu) (1 sfu = 10$^4$ Jansky (Jy), where
1 Jy = 10$^{-26}$ Wm$^{-2}$ Hz$^{-1}$). The solar radio flux data, recorded 
by four sites around the globe, are made available by the NOAA SWPC
(\url{https://www.ngdc.noaa.gov/stp/space-weather/solar-data/}). 
In Figure \ref{fig_15}, a gradual increase in the solar activity can been seen in 
most of the flux density profiles, between 13 December 2021 and 10 April 2023. 

\subsection{Radio Flux Density and Brightness Temperature}

\begin{figure}[t]
  \centerline{\hspace{0.0\textwidth}
              \includegraphics[width=0.7\textwidth,angle=-90]{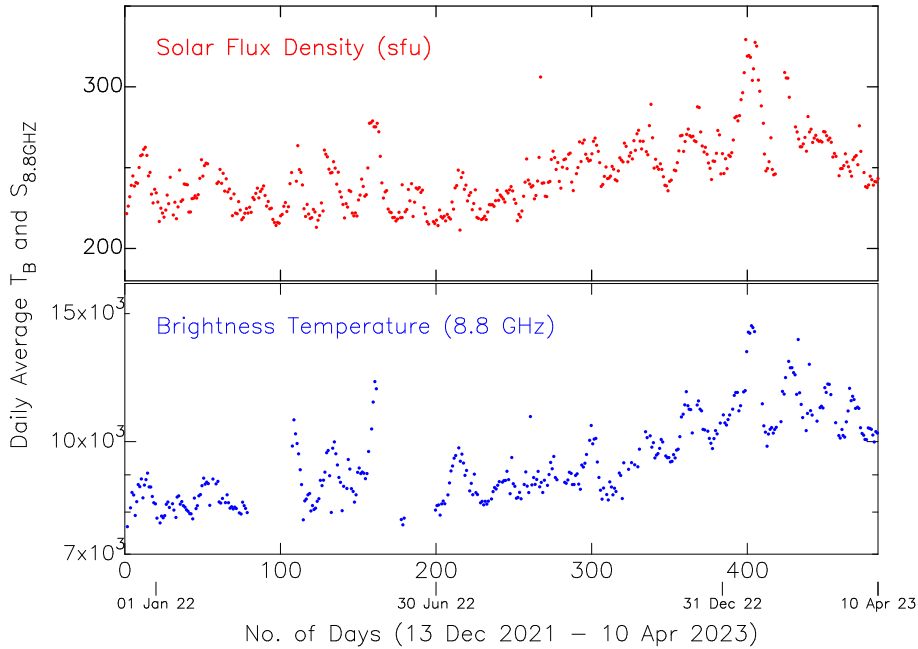}
              }
\caption{(Upper) Daily average solar radio flux density observed with the 2.4-m Learmonth 
Solar Observatory telescope, which is part of the RSTN network, at 8.8 GHz plotted between 
13 December and 09 April 2023.  (Lower) Daily average brightness temperature observed at 
8.8 GHz with the Arecibo 12-m telescope. The variations seen in the radio flux density 
profile agree well with those seen in the average Arecibo brightness temperature.  The 
overall increasing trend seen in both profiles is in agreement with the increasing solar 
activity from the start to end of the data stretch.
}
\label{fig_16}
\end{figure}

It will be effective to compare the solar flux density observed at 8.8 GHz at the 
Learmonth Solar Observatory (Figure \ref{fig_15}) with the Arecibo daily average 
radio brightness temperature at the same frequency. This
provides a relationship between the brightness temperature and solar flux density
measured by two independent telescopes.
Figure \ref{fig_16} (top panel) shows the daily average flux density measured at 
8.8 GHz by the 2.4-m telescope at Learmonth Solar Observatory for the period
between 13 December 2021 and 09 April 2023 (the flux-density profile at 8.8 
GHz is reproduced from Figure \ref{fig_15}). 
In the bottom panel of the figure, the daily average brightness temperature at 
8.8 GHz from the Arecibo 12-m telescope is plotted. The detailed profiles from 
the two telescopes show one-to-one correlation.
In particular, the overall trend of increasing intensity of the emission from the start 
of the profiles to the end, in association with the increasing solar activity, also 
shows the agreement between these two observations. 

\begin{figure}[t]
  \centerline{\hspace{0.05\textwidth}
              \includegraphics[width=0.6\textwidth,angle=-90]{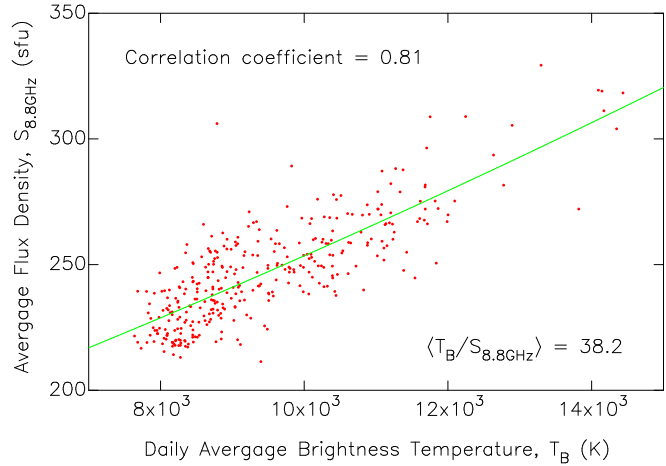}
              }
\caption{The correlation plot between the daily average solar flux density measurements
from the Learmonth Solar Observatory (RSTN) and the brightness temperature estimates
from the 12-m telescope at 8.8 GHz.  The linear correlation coefficient between these 
two parameters is 0.81.  The average ratio between the brightness temperature and the 
flux density is 38.2~K/sfu.
}
\label{fig_17}
\end{figure}

Figure \ref{fig_17} shows the correlation plot between the daily average flux density 
at Learmonth and the daily average Arecibo brightness temperature at 8.8 GHz. 
(In this plot, at brightness temperatures $>$10000~K, the flux densities tend to deviate 
moderately towards lower values, which may require a through investigation 
of the linearity of the flux density measurements at high levels of activity.)
For a given brightness temperature, mean-to-peak fluctuations of $\leq$10\% in the radio 
flux densities are observed. An overall linear correlation coefficient of 81\% between the 
two measured parameters indicates the close agreement between them. 
This correlation was further checked and confirmed, in the Rayleigh-Jeans approximation 
to Planck$'$s radiation law, by converting the observed X-band brightness temperature to 
flux density over the disk of the Sun.
In this phase of solar cycle \#25, the
average flux density at 8.8 GHz was $\sim$245 sfu. The daily ratios between the  
brightness temperature and the flux density at 8.8 GHz, (T$_{\rm B}$/S$_{8.8{\rm GHz}}$), 
vary between $\sim$28 and $\sim$50 and have an average value of 38.2 K$/$sfu.  

\section{Concluding Remarks}

In this paper, we present X-band solar mapping observations made with the Arecibo 12-m radio 
telescope in the frequency range of 8.1 -- 9.2 GHz. These observations have revealed the 
highly complex and variable brightness distribution of solar features over several solar
rotations of the current solar cycle \#25, covering a fair portion of the ascending phase
between 13 December 2021 and 09 April 2023. The solar maps have been used to
locate and track active regions of space-weather importance.  The radio signatures 
of an active region in the frequency range of 8 -- 9 GHz provide a powerful diagnostic of 
the gyro-synchrotron radiation from high-energy electrons trapped in small-scale magnetic 
field loops (e.g., \citealt{nindos2020}) and the temporal brightness temperature changes of 
active regions have been 
compared with their area and magnetic configuration. Although a correlating tendency 
has been observed between the brightness temperature and the area of the active region,
(also with the number of sunspots within it), understanding the differences in the formation, 
development, and decay of an active region in terms of its area and/or magnetic configuration 
alone poses a great challenge. Thus, a given area of an active group, (or the number of spots 
within it), may not be a precise parameter when deciding on the actual capacity of an eruption 
site. The results of the present study demonstrates that the X-band solar mapping provides an
improved 
picture of the formation location of an active region and helps to accurately 
track its evolution across the solar disk, forewarning of intense solar eruptions
leading to severe space-weather consequences.

Additionally, the ``Latitude -- Time'' plot over several solar rotations provides the
evolution of quiet regions on the Sun, coronal holes, and eruptive sites. This also
provides the three-dimensional picture of radio emission 
(i) above sunspots, where the magnetic field is stronger, but less divergent, (ii) 
high- and mid-latitude coronal holes of largely unipolar and open field regions, and
(iii) large-scale emitting structures, 
which can be linked to the flux
emergence rate from regions below the photosphere. The present analysis emphasizes 
the importance of the long-term monitoring of the Sun at X-band for understanding the 
complex three-dimensional evolution of solar features as a function of solar activity.
The agreement obtained between the daily radio flux density and brightness temperature
data provides a typical scaling factor involved. However, this has to be further
thoroughly checked for slowly varying emission and sporadic bursts caused by flares
and CMEs with improved spatial resolution measurements. 


During the X-band solar mapping observations, several CME events were also detected. 
These observations are useful towards understanding hot magnetized plasma conditions
at the initial stages of CMEs at the chromospheric height. For academic interest, we 
also tracked a few active regions and recorded the brightness temperatures at the high 
temporal resolution. Results of the CME-event study and emission profile analysis will 
be presented elsewhere.
  

The 12-m radio telescope is currently being upgraded with a wideband, 2.3 -- 14 GHz, cooled
front-end system, which will considerably enhance its sensitivity, as well as its frequency
coverage. Since the new receiver also allows measurement of full-Stokes parameters, its 
extended bandwidth will provide highly accurate temporal measurements of polarization and 
dynamic spectra of the solar emission. This will be valuable for studying the evolution of 
the magnetic-field configurations and plasma conditions in the magnetic current sheets of 
solar eruptions, which are essentially required for understanding the origin of space-weather 
events. Moreover, the mapping of the Sun over a wide frequency band will also provide the 
temporal and spatial evolution of the eruptive active regions at different layers between 
the photosphere and the lower corona.

In addition, the upgraded 12-m system will allow interplanetary scintillation (IPS) 
observations of compact background radio sources that can probe the ambient solar wind and 
structures within propagating CMEs in the three-dimensional inner heliosphere, for regions 
inaccessible to spacecraft, between the solar wind acceleration region and to about the 
middle Sun-Earth distance ($\sim$10 -- 100 solar radii) (e.g., \citealt{mano2010}). Indeed 
the wide bandwidth made available will allow us to probe solar-wind density structure of 
different scale sizes, {\it i.e.}, Fresnel radii, which are useful towards understanding the 
plasma properties associated with propagating space-weather events (e.g., \citealt{cohen1969}; 
\citealt{mano1993}).  A set of solar observations at high- and low-radio frequencies 
respectively with the Arecibo 12-m radio telescope and the Arecibo Callisto Radio Spectrometer 
(\url{https://www.e-callisto.org/}),
combined with IPS measurements, will be an asset for a detailed understanding of space-weather 
events in the Sun-Earth space.

\begin{acks}

The Arecibo Observatory is operated by the University of Central Florida
under a cooperative agreement with the National Science Foundation
(AST-1822073), and in alliance with Universidad Ana G. M{\'e}ndez and Yang 
Enterprises, Inc. PKM wishes to thank Tapasi Ghosh for the numerous useful 
discussions and suggestions during the stages of analysis and the preparation 
of the manuscript. We acknowledge the EUV data from the UVE and images from 
the AIA and HMI on board the Solar Dynamics Observatory. The X-ray data sets 
have been obtained from the Geostationary Operational Environmental Satellite 
(GOES-16). We also acknowledge OMNIdata of NASA/GSFC’s Space Physics Data 
Facility. The Wilcox Solar Observatory provided the source surface magnetic 
field data. The SOHO/LASCO is a project of international cooperation between 
ESA and NASA. We are grateful to the Learmonth Solar Observatory, one of the 
stations of the Radio Solar Telescope Network (RSTN), for the provision of 
the solar flux density data.

\end{acks}

\bibliographystyle{spr-mp-sola}
\bibliography{manuscript}

\end{article} 
\end{document}